\documentclass[aps,twocolumn,showpacs,floats]{revtex4}

\usepackage{graphicx}

\begin{document}

\title{Renormalization group dependence of the QCD coupling}

\author{G.\ X.\ Peng%
\footnote{Email: gxpeng@ihep.ac.cn, gxpeng@lns.mit.edu}
       }

\affiliation{
 $^{\mathrm{a}}$China Center of Advanced Science and Technology
     (World Lab.), P.O.Box 8730, Beijing 100080, China\\
 $^{\mathrm{b}}$Institute of High Energy Physics, Chinese Academy of Sciences,
     P.O.Box 918, Beijing 100039, China\\
 $^{\mathrm{c}}$Center for Theoretical Physics MIT,
    77 Massachusetts Avenue, Cambridge, MA 02139-4307, USA
             }


\begin{abstract}
The general relation between the standard expansion coefficients and
the beta function for the QCD coupling is exactly derived in a
mathematically strict way. It is accordingly found that an infinite
number of logarithmic terms are lost in the standard expansion with
a finite order, and these lost terms can be given in a closed form.
Numerical calculations, by a new matching-invariant coupling with
the corresponding beta function to four-loop level, show that the
new expansion converges much faster.
\end{abstract}


\maketitle

It is of crucial importance to consider the renormalization
group (RG) scale dependence of the strong coupling, in order
to have full consistency in QCD and its applications
\cite{Peng2005epl,Fraga2005prd71}. As is well known, the QCD
running coupling $\alpha=\alpha_s/\pi=g^2/(4\pi^2)$ satisfies
the RG equation
\begin{equation} \label{RGa}
u\frac{\mathrm{d}\alpha}{\mathrm{d}u}
=-\sum_{i=0}^{\infty}
   \beta_i\alpha^{i+2}
\equiv \beta(\alpha),
\end{equation}
where the $\beta$\ function
was calculated to one-loop level in QCD more than thirty
years ago \cite{Gross1973PRL30.1343},
to two-loop in Ref.\ \cite{Caswell1974PRL33.244},
to three-loop in Ref.\ \cite{Tarasov1980PLB93.429},
and to four-loop in Ref.\ \cite{Larin1997PLB400.379},
in the minimal subtraction scheme \cite{tHooft1973NPB61.455}.
It can be expressed as polynomials of the number of flavors
 $N_{\mathrm{f}}$, i.e.,
$
\beta_i(N_{\mathrm{f}})=\sum_{j\geq 0} \beta_{i,j}
N_{\mathrm{f}}^j.\
$
The color factor $\beta_{i,j}$ is presently available to
4-loop level, i.e.,
\begin{equation}
 [\beta_{i,j}]=\left[
  \begin{array}{cccc}
  11/2  &  -1/3    &    0     &   0 \\
  51/4  & -19/12   &    0     &   0 \\
\frac{2857}{64} &-\frac{5032}{576} & \frac{325}{1728} &   0 \\
 \beta_{3,0} &   \beta_{3,1} &  \beta_{3,2}  & \frac{1093}{93312} \\
  \end{array}
          \right]
\end{equation}
with
$
\beta_{3,0}=891\zeta_3/32+149753/768 \approx 228.4606573, \
\beta_{3,1}=-1627\zeta_3/864-1078361/20736 \approx -54.26788763, \
\beta_{3,2}=809\zeta_3/1296+50065/20736 \approx 3.164758128,
$
where $\zeta$\ is the Riemann zeta function,
and
$\zeta_2=\pi^2/6$,
$\zeta_3\approx 1.202056903$,
$\zeta_4=\pi^4/90$,
$\zeta_5\approx 1.036927755$.
In this paper, all color factors are given for $N_{\mathrm{c}}=3$.
Comparing the beta expressions here with those
in Ref.\ \cite{Larin1997PLB400.379},
one would find a difference by a factor of $2^{2i+1}$.

In practical applications, it is convenient to have an explicit expression
of the $\alpha$\ as a function of the renormalization point $u$.
The standard approach is to expand it to a series of
$L=1/\ln(u^2/\Lambda^2)$, where $\Lambda$\ is the QCD
scale parameter. However, how the expansion coefficients
are connected to the beta function is not generally known, though
one can find the relation to order 3 in Ref.\ \cite{Eidelman2004PLB592.1},
and to order 4 in Ref.\ \cite{Chetyrkin1997PRL79}.
In this letter, the general relation between the expansion
coefficients and the beta function are provided.
It is accordingly found that an infinite number of terms like $\ln^jL$ are
lost in the standard expansion with a finite order, and these lost
terms can be given in a closed form.
Numerical calculations, by a new matching-invariant
coupling with the corresponding beta function to four-loop level,
show that the new expansion, with the lost terms included,
converges much faster.

To solve Eq.\ (\ref{RGa}), let's define
\begin{equation}
\acute{\beta}(\alpha)
\equiv
  \frac{\beta_0^2}{\beta_1}
  \left[
    \frac{1}{\beta(\alpha)}
   -\frac{\beta_1}{\beta_0^2\alpha}
   +\frac{1}{\beta_0\alpha^2}
  \right].
\end{equation}
With the series expression for $\beta(\alpha)$ in Eq.\ (\ref{RGa}),
one can easily get an explicit expression
\begin{equation} \label{abeta2}
\acute{\beta}(\alpha)
=
 \frac{\sum_{j=0}^{\infty}(\beta_0\beta_{j+2}/\beta_1
                           -\beta_{j+1})\alpha^j}
  {\sum_{i=0}^{\infty}\beta_i\alpha^i},
\end{equation}
which indicates that $\acute{\beta}(\alpha)$ is analytic
at $\alpha=0$, and can thus be expanded to a Taylor series
as
\begin{equation} \label{abeta3}
\acute{\beta}(\alpha)
= \sum_{k=0}^{\infty} \acute{\beta}_k\alpha^k.
\end{equation}
The expansion coefficients $\acute{\beta}_k$ can be obtained
from the normal mathematical formula
\begin{equation} \label{aktilde}
\acute{\beta}_k
=\frac{1}{k!}
 \left.
 \frac{\mathrm{d}^k}{\mathrm{d}\alpha^k} \acute{\beta}(\alpha)
 \right|_{\alpha=0}.
\end{equation}
Another easy way to obtain these coefficients is
to use the recursive relation
\begin{equation} \label{grbetarec}
\acute{\beta}_k
= \frac{\beta_{k+2}}{\beta_1}
 -\frac{\beta_{k+1}}{\beta_0}
 -\frac{1}{\beta_0}\sum_{l=0}^{k-1}\beta_{k-l}\acute{\beta}_l,
\end{equation}
with the obvious initial condition
$ 
\acute{\beta}_0
={\beta_2}/{\beta_1}-{\beta_1}/{\beta_0}.
$ 

Let
$ 
L=1/\ln(u/\Lambda),
$ 
where $\Lambda$\ is a dimensional parameter,
then Eq.\ (\ref{RGa}) becomes
$ 
-L^2\frac{\mathrm{d}\alpha}{\mathrm{d}L}
=\beta(\alpha),
$ 
or
$ 
-\frac{\mathrm{d}L}{L^2}
=\frac{\mathrm{d}\alpha}{\beta(\alpha)}
=\left[
  -\frac{1}{\beta_0\alpha^2}
  +\frac{\beta_1}{\beta_0^2\alpha}
  +\frac{\beta_1}{\beta_0^2}\acute{\beta}(\alpha)
 \right]
 \mathrm{d}\alpha.
$ 
Integrating this equation gives
$ 
\frac{1}{L}-C'
=\frac{1}{\beta_0\alpha}
 +\frac{\beta_1}{\beta_0^2}\ln\alpha
 +\frac{\beta_1}{\beta_0^2}
  \int_0^{\alpha} \acute{\beta}(x) \mathrm{d} x,
$ 
or
\begin{equation}
 1-C' L
=\frac{L}{\beta_0\alpha}
 +\frac{\beta_1}{\beta_0^2}L\ln\alpha
 +\frac{\beta_1 L}{\beta_0^2}
  \sum_{k=1}^{\infty}\frac{\acute{\beta}_{k-1}}{k}\alpha^k,
\end{equation}
where $C'$ is the constant of integration.

Let
$ 
\alpha=\frac{L}{\beta_0}Y(L),
$ 
then
\begin{equation} \label{eqY}
\frac{1}{Y}
 +L^*\ln Y
 +\sum_{k=1}^{\infty}\grave{\beta}_k {L^*}^{k+1}Y^k
=1-CL^*-L^*\ln L,
\end{equation}
where
$L^*\equiv (\beta_1/\beta_0^2)L$,
$C\equiv (\beta_0^2/\beta_1) C'-\ln\beta_0$,
and
\begin{equation}
\grave{\beta}_k
\equiv
\frac{1}{k}
\left(\frac{\beta_0}{\beta_1}\right)^k\acute{\beta}_{k-1}.
\end{equation}
The graved beta function $\grave{\beta}_k$
can be easily obtained from the expression for
the acute beta function $\acute{\beta}_k$ in Eq.\ (\ref{aktilde})
or (\ref{grbetarec}), and here are the results:
\begin{eqnarray}
\grave{\beta}_1
 &=& -1+\dot{\beta}_0\dot{\beta}_2, \
\grave{\beta}_2
 = 1/2-\dot{\beta}_0\dot{\beta}_2+(1/2){\dot{\beta}}_0^2\dot{\beta}_3, \\
\grave{\beta}_3
 &=&
  -\frac{1}{3}+\dot{\beta}_0\dot{\beta}_2
  -\frac{1}{3}{\dot{\beta}}_0^2(2\dot{\beta}_1\dot{\beta}_3+{\dot{\beta}}_2^2)
  +\frac{1}{3}\dot{\beta}_0^3\dot{\beta}_4, \\
\grave{\beta}_{k\ge 4}
 &=&\frac{(-1)^k}{k}+(-1)^{k-1}\dot{\beta}_0\dot{\beta}_2
 +\frac{1}{k}\sum_{s=1}^{k-4} \Bigg[
 (-1)^s {\dot{\beta}_0}^{k-s-2}
  \nonumber\\
&& \hspace{-0.5cm}
  \times \sum_{r=0}^2
 \left(
  \prod_{p=1}^{s+r}\sum_{l_p=s+r-p+1}^{l_{p-1}-1}
 \right)
  B_{l,s}^{(r)}
 \prod_{q=0}^{s+r-1}\beta_{l_q-l_{q+1}} \Bigg]
 \nonumber\\
&&  \hspace{-0.5cm}
 +\frac{{\dot{\beta}_0}^{k-2}}{k}
 \Bigg[
  \dot{\beta}_{l_0}
  +\sum_{l_1=1}^{k-2}
   (\dot{\beta}_{l_1+1}+\dot{\beta}_2\dot{\beta}_{l_1})\dot{\beta}_{l_0-l_1}
 \nonumber\\
 & &  \hspace{-0.5cm} \phantom{+\frac{{\dot{\beta}_0}^{k-2}}{k}}
  +\sum_{l_1=2}^{k-2}\sum_{l_2=1}^{l_1-1}
   \dot{\beta}_{l_0-l_1}\dot{\beta}_{l_1-l_2}\dot{\beta}_{l_2+2}
  \Bigg]
\nonumber\\
&& \hspace{-0.5cm}
 -\frac{{\dot{\beta}_0}^{k-1}}{k}
 \left(
  \dot{\beta}_k+\sum_{s=0}^{k-2}\dot{\beta}_{l_0-s}\dot{\beta}_{s+2}
 \right)
 +\frac{{\dot{\beta}_0}^k\dot{\beta}_{k+1}}{k},
\end{eqnarray}
where
  $l_0=k-1$,
 $B_{l,s}^{(0)}=\dot{\beta}_{l_s}$,
 $B_{l,s}^{(1)}=\dot{\beta}_{l_{s+1}+1}+\dot{\beta}_2\dot{\beta}_{l_{s+1}}$,
 $B_{l,s}^{(2)}=\dot{\beta}_{l_{s+2}+2}$,
and $\dot{\beta}_i=\beta_i/\beta_1$ ($i=0, 1, 2, \ldots$).

Now suppose we have a solution of the form
\begin{equation}
Y(L)
=\sum_{i=0}^{\infty}\sum_{j=0}^{i}
 f_{i,j} {L^*}^i\ln^j L,
\end{equation}
then
\begin{equation}
\frac{1}{Y}
=\frac{1}{f_{0,0}}+\sum_{i=1}^{\infty}\sum_{j=0}^{i}
 \left(\sum_{k=1}^{i}\frac{(-1)^k}{f_{0,0}^{k+1}} \bigsqcup_{1,0}^k f_{i,j}\right)
 {L^*}^i\ln^j L,
\end{equation}
\begin{equation}
L^*\ln Y
= 
 \sum_{i=2}^{\infty}\sum_{j=0}^{i-1}
 \left(\sum_{k=1}^{i-1}\frac{(-1)^{k-1}}{kf_{0,0}^k}
  \bigsqcup_{1,0}^k f_{i-1,j}\right){L^*}^i\ln^j L,
\end{equation}
\begin{equation}
\sum_{k=1}^{\infty} \grave{\beta}_k{L^*}^{k+1}Y^k
=\sum_{i=2}^{\infty}\sum_{j=0}^{i-2}
 \left(
  \sum_{k=2}^{i-j}\grave{\beta}_{k-1} \bigsqcup_{0,0}^{l-k} f_{i-k,j}
 \right)
 {L^*}^i\ln^j L,
\end{equation}
where the square cup operator, $\bigsqcup$, has been defined
in the appendix.
Substituting these expressions into Eq.\ (\ref{eqY}),
we can obtain all $f_{i,j}$ by comparing
the corresponding coefficients of ${u^*}^i\ln^ju$.
For $(i,j)=(0,0)$, $(1,0)$, and $(1,1)$,
we have
$ 
f_{0,0}=1, \ f_{1,0}=C, \ f_{1,1}=1.
$ 
%
For $i\geq 2$ and $j=i$, we have
$\sum_{k=1}^i (-1)^k \bigsqcup_{1,0}^k f_{i,i}=0$,
which gives $f_{i,i}=1$. And for $i\geq 2$ and $j=i-1$, we get
\begin{equation}
\sum_{k=1}^i (-1)^k \bigsqcup_{1,0}^k f_{i,i-1}
+\sum_{k=1}^{i-1}\frac{(-1)^{k-1}}{k}\bigsqcup_{1,0}^k f_{i-1,i-1}
=0
\end{equation}
whose solution is
$
f_{i,i-1}
= iC+\sum_{l=1}^{i-1}\left({i}/{l}-1\right).
$

For $i\geq 2$ and $0\leq j \leq i-2$, we have
\begin{eqnarray} \label{eqfij3}
f_{i,j}
 &=&
 \sum_{k=2}^i (-1)^k \bigsqcup_{1,0}^k f_{i,j}
+\sum_{k=1}^{i-1}\frac{(-1)^{k-1}}{k}\bigsqcup_{1,0}^k f_{i-1,j}
 \nonumber\\
&&
+\sum_{k=1}^{i-j-1}\grave{\beta}_k\bigsqcup_{0,0}^kf_{i-k-1,j}.
\end{eqnarray}
Please note, there are only terms of $f_{i'<i,j'<j}$
 on the right hand side of this equation.
Therefore, it is a recursive relation.
Here are the solution to order 5:
\begin{eqnarray}
f_{2,0} &=& C^2+C+\grave{\beta}_1,  \\
f_{3,0} &=& C^3+\frac{5}{2}C^2+(3\grave{\beta}_1+1)C
            +\grave{\beta}_1+\grave{\beta}_2,\\
f_{3,1} &=& 3C^2+5C+3\grave{\beta}_1+1, \\
f_{4,0} &=& C^4+(13/3)C^3+(6\grave{\beta}_1+9/2)C^2
     \nonumber\\
        & &
     +(1+7A_1+4\grave{\beta}_2)C+2\grave{\beta}_1^2+\Sigma_{i=1}^{3}\grave{\beta}_i, \\
f_{4,1} &=& 4C^3+13C^2+(12\grave{\beta}_1+9)C
     \nonumber\\
        &&
            +7A_1+4\grave{\beta}_2+1, \\
f_{4,2} &=& 6C^2+13C+6\grave{\beta}_1+9/2, \\
\cdots \nonumber
\end{eqnarray}

These correspond to the standard form in Ref.\ \cite{Eidelman2004PLB592.1}
at order 3, and agrees to that in Ref.\ \cite{Chetyrkin1997PRL79}
at order 4, i.e.,
\begin{eqnarray}
\alpha(u)
&=&\frac{1}{\beta_0\ln(u/\Lambda)}
 \left\{
  1-\frac{\beta_1\ln\ln(u/\Lambda)}
         {\beta_0^2\ln(u/\Lambda)}
\right.\nonumber\\
&&  \hspace{-0.5cm}
 \left.
 +\frac{\beta_1^2}{\beta_0^4\ln^2(u/\Lambda)}
 \left[
  \left(\ln\ln\frac{u}{\Lambda}
                                 -\frac{1}{2}\right)^2
  +\frac{\beta_0\beta_2}{\beta_1^2}
  -\frac{5}{4}
 \right]
\right.\nonumber\\
&& \hspace{-0.5cm}
 \left.
 -\frac{\beta_1^3}{\beta_0^6\ln^3(u/\Lambda)}
 \left[
 \left(\ln\ln\frac{u}{\Lambda}-\frac{5}{6}\right)^3
  -\frac{\beta_0^2\beta_3}{2\beta_1^3}
  +\frac{233}{216}
\right.\right.\nonumber\\
&&  \hspace{-0.5cm}
 \phantom{-\frac{\beta_1^3}{\beta_0^6\ln^3(u/\Lambda)}[}
  +\left(3\frac{\beta_0b_2}{\beta_1^2}-\frac{49}{12}\right)
   \ln\ln\frac{u}{\Lambda}
 \Bigg]
 \Bigg\}.
\end{eqnarray}

To give a general representation for the expansion coefficients $f_{i,j}$,
introduce a set of new functions $\bar{\beta}_i$ by
the recursive relation
\begin{eqnarray} \label{eqbetabar}
\bar{\beta}_i
=\sum_{k=1}^{i-1}
 \left[
  K
  \left(\bigsqcup_1^{k+1} \bar{\beta}_i
  -\bigsqcup_1^k \bar{\beta}_{i-1}\right)
 \right.
   \left.
  +\grave{\beta}_k\bigsqcup_0^k \bar{\beta}_{i-k-1}
 \right],
\end{eqnarray}
where
$K\equiv (-1)^{k-1}(1+1/k)$.
From the initial conditions $\bar{\beta}_0=1$ and $\bar{\beta}_1=C$,
one can easily get all $\bar{\beta}_i$ from  Eq.\ (\ref{eqbetabar}).
For $C=0$, for example, we have
\begin{eqnarray}
\bar{\beta}_0
 &=& 1, \
\bar{\beta}_1
 = 0, \
\bar{\beta}_2
 =\grave{\beta}_1, \
\bar{\beta}_3
 = \grave{\beta}_1+\grave{\beta}_2, \\
\bar{\beta}_4
 &=&
  \sum_{i=1}^3\grave{\beta}_i+2\grave{\beta}_1^2, \
\bar{\beta}_5
= \sum_{i=1}^4\grave{\beta}_i +\frac{9}{2}\grave{\beta}_1^2
   +5\grave{\beta}_1\grave{\beta}_2, \\
\bar{\beta}_6
 &=&
 \Sigma_{i=1}^5\grave{\beta}_i +\frac{15}{2}\grave{\beta}_1^2
 +3\grave{\beta}_2^2
 +11\grave{\beta}_1\grave{\beta}_2
 +5\grave{\beta}_1^3, \\
 &\cdots& \nonumber
\end{eqnarray}
Please note, even when one sets all $\beta_{i>2}$ to zero,
no $\bar{\beta}_{k>2}$ will be zero. This is one of the most
important reason for us to know the general relation between
$f_{i,j}$ and the beta function.

It is found that $f_{i,j}$ satisfies
\begin{equation} \label{fijrecur}
f_{i,j}
 = \sum_{k=0}^{i-j} f_{i-j,k}
   \sum_{l=0}^j
   \frac{(i-l)!}{(i-j)!(j-l)!}
    \bigsqcup_0^k \Phi_l,
\end{equation}
where
$\bigsqcup_0^k \Phi_l$
are the Taylor coefficients of $\ln^k(1+x)$, i.e.,
$
\ln^k(1+x)
=\sum_{l=0}^{\infty} \left(\bigsqcup_0^k\Phi_l\right)x^l
=\left(\sum_{l=0}^{\infty}\Phi_l x^l\right)^k
=\left(\sum_{l=1}^{\infty}\frac{1}{l}x^l\right)^k.
$
 Accordingly we have
$
\Phi_0=0,\
\Phi_{l\ge 1}=1/l,\
\bigsqcup_0^0\Phi_l=\delta_{l,0},\
\bigsqcup_0^k\Phi_0=\delta_{k,0},\
\bigsqcup_0^1\Phi_l=\Phi_l,
$
 and 
\begin{equation}
\bigsqcup_0^{k\ge 2} \Phi_{l\ge 1}
=\left(
  \prod_{s=1}^{k-1}
  \sum_{p_s=1}^{l-k+s-\Sigma_{q=1}^{k-1} p_q }
 \right)
   \frac{1}{l-\Sigma_{q=1}^{k-1}p_q}
   \prod_{s=1}^{k-1} \frac{1}{p_s}.
\end{equation}

One can naturally consider to prove Eq.\ (\ref{fijrecur})
by mathematical induction. We have numerically checked it to order 100.
Now we directly used it to give an expression for $f_{i,j}$, i.e.,
\begin{equation} \label{figexpgen}
f_{i,j}=\sum_{k=0}^{i-j} P_{j,k}^{(i)} \bar{\beta}_k,
\end{equation}
where $P_{j,k}^{(i)}$ can be obtained by repeatedly
using Eq.\ (\ref{fijrecur}). For $k=i-j$ and $k=i-j-1$,
it is easy to get
\begin{equation}
P_{j,i-j}^{(i)}
=\frac{i!}{j!(i-j)!}, \
P_{j,i-j-1}^{(i)}
=\frac{i-j}{(i-j)!}
 \sum_{l=1}^j\frac{(i-l)!}{(j-l)!l}.
\end{equation}
 Generally for $k=i-j-l$ or $l=i-j-k$, careful derivations give
\begin{eqnarray}
P_{j,k}^{(i)}
&=& \mathcal{Q}(i,j,l) \mathcal{Q}(i-j,l,0)
\nonumber\\
&&
   +\sum_{r=1}^{l-1} \Bigg\{
    \left(
     \prod_{p=1}^r\, \sum_{s_p=1}^{l-1-\Sigma_{q=1}^{p-1}s_q}
    \right)
  \Big[ \mathcal{Q} (i,j,s_1)
\nonumber\\
&& \times
  \mathcal{Q}\left(i-j-\Sigma_{q=1}^{r-1}s_q,s_r,l-\Sigma_{q=1}^{r}s_q\right)
\nonumber\\
&&\times
  \mathcal{Q}\left(i-j-\Sigma_{q=1}^r s_q,l-\Sigma_{q=1}^r s_q,0\right)
\nonumber\\
&&\times
  \prod_{p=1}^{r-1} \mathcal{Q}\left(i-j-\Sigma_{q=1}^{p-1}s_q,s_p,s_{p+1}\right)
  \Big] \Bigg\},
\end{eqnarray}
where $\mathcal{Q}$ is a function of three non-negative
integers and defined by
\begin{equation}
\mathcal{Q}(i,j,k)
\equiv
\frac{1}{(i-j)!}
\sum_{l=0}^j
\frac{(i-l)!}{(j-l)!} \bigsqcup_0^k\Phi_l.
\end{equation}

For a given $i$, one can regard $P^{(i)}=\left[P_{j,k}^{(i)}\right]$
as a matrix of order $i+1$.
The specially simple elements are:
$ P_{i,k\ge i-j}^{(i)}=0, \
 P_{0,k}^{(i)}=\delta_{k,i}, \
 P_{1,k}^{(i)}=k+1, \
 P_{i,k}^{(i)}=\delta_{k,0}, \
 P_{i-1,0}^{(i)}
 =\sum_{s=1}^{i-1}(i/s-1), \
P_{i-1,1}^{(i)}=i.
$\
Here are the $P^{(i)}$ matrix for $i$ from 0 to 4:
\begin{eqnarray}
P^{(0)}=[1],
\
P^{(1)}=
  \left[
   \begin{array}{cc}
    0 & 1 \\
    1 & 0
   \end{array}
  \right],
\
P^{(2)}=
  \left[
   \begin{array}{ccc}
    0 & 0 & 1 \\
    1 & 2 & 0 \\
    1 & 0 & 0
   \end{array}
  \right],
\nonumber\\
P^{(3)}=
  \left[
   \begin{array}{cccc}
    0   & 0 & 0 & 1 \\
    1   & 2 & 3 & 0 \\
    \frac{5}{2} & 3 & 0 & 0 \\
    1   & 0 & 0 & 0
   \end{array}
  \right],
\
P^{(4)}=
  \left[
   \begin{array}{ccccc}
    0 &  0   & 0 & 0 & 1 \\
    1    & 2 & 3 & 4 & 0 \\
    {9}/{2}  & 7 & 6 & 0 & 0 \\
    {13}/{3} & 4 & 0 & 0 & 0 \\
    1    & 0 & 0 & 0 & 0
   \end{array}
  \right].
\end{eqnarray}

In the traditional minimum subtraction scheme ($\overline{\mathrm{MS}}$),
the strong coupling $\alpha(u)$ as a function of the
re-normalization point $u$\ is not continuous at the
quark masses. Let's derive a matching-invariant coupling
by absorbing loop effects into the $\overline{\mathrm{MS}}$
definition and give the corresponding beta function to
four-loop level.

Suppose the new coupling $\alpha'$ is connected to the original
coupling $\alpha$ by
\begin{equation} \label{apdef}
\alpha'
=\sum_{i=0}^{\infty}a_i \alpha^{i+1}.
\end{equation}
Then, using the matching condition
$
\check{\alpha}
=\sum_{j=0}^{\infty} C_j \alpha^{j+1}
$
with the matching coefficients \cite{Chetyrkin1997PRL79,Rodrigo1993PLB313}
\begin{eqnarray}
C_0 &=& 1, \ \  C_1 = 0, \ \ C_2 = 11/72, \\
C_3  \label{C30exp}
&=&
 \frac{575263}{124416}
 -\frac{82043}{27648}\zeta_3
 -\frac{2633}{31104}N_{\mathrm{f}}, \
\cdots
\end{eqnarray}
one has
\begin{equation}
\check{\alpha'}
=\sum_{i=0}^{\infty}
 \left[
  \sum_{k=0}^i
  \check{a}_k\bigsqcup_0^{k+1}C_{i-k}
 \right]
 \alpha^{i+1},
\end{equation}
where an overhead check means decreasing $N_{\mathrm{f}}$
by one flavor to the corresponding ($N_{\mathrm{f}}-1$)-flavor effective
theory. Accordingly, comparing the coefficients of $\alpha$
in the equality $\check{\alpha}^{\prime}=\alpha^{\prime}$ yields
\begin{equation} \label{aisqcup}
a_i
=\sum_{k=0}^i
 \check{a}_k\bigsqcup_0^{k+1}C_{i-k}.
\end{equation}
Assume
$ 
a_i
=\sum_{j=0}^i a_{i,j}N_{\mathrm{f}}^j,
$ 
then $\check{a}_k=\sum_{j=0}^k a_{k,j}(N_{\mathrm{f}}-1)^j$.
Substitution into Eq.\ (\ref{aisqcup}) then gives
\begin{equation}
\sum_{k=0}^i
\left[
a_{i,k}N_{\mathrm{f}}^k
-\left(\bigsqcup_0^{k+1}C_{i-k}\right)
 \sum_{j=0}^k a_{k,j}(N_{\mathrm{f}}-1)^j
\right]=0,
\end{equation}
whose solution is
\begin{eqnarray}
a_0 &=& a_{0,0}, \ \
a_1 = a_{1,0},  \ \
a_2 = a_{2,0}+\frac{11}{72}a_{0,0}N_{\mathrm{f}}, \\
a_3 &=& a_{3,0}
 +\left[
    \left(\frac{7037}{1536}
     +\frac{82043}{27648}\zeta_3
    \right)a_{0,0}
   +\frac{11}{36}a_{1,0}
  \right] N_{\mathrm{f}}
\nonumber\\
&&
  -\frac{2633}{62208}a_{0,0}N_{\mathrm{f}}^2, \ \  \cdots 
\end{eqnarray}
To definitely fix the new coupling, one needs to choose $a_{i,0}$.
The simplest choice would be
$a_{i,0}=\delta_{i,0}.$\
 With this convention, one has
\begin{eqnarray}
&& a_0=1, \ \ a_1=0, \ \ a_2=(11/72)N_{\mathrm{f}},
               \label{a012exp}  \\
&& a_3
    =a_{3,1}
     N_{\mathrm{f}}
    -\frac{2633}{62208}N_{\mathrm{f}}^2, \ \ \cdots
 \label{a3exp}
\end{eqnarray}
where
$
a_{3,1}=7037/1536+82043\zeta_3/27648
\approx 8.148377983.
$
Then the new matching-invariant coupling is
\begin{equation}
\alpha^{\prime}
=
 \alpha
 +\frac{11}{72}N_{\mathrm{f}}\alpha^3
 +\left(
   a_{3,1}
   -\frac{2633}{62208}N_{\mathrm{f}}
  \right)N_{\mathrm{f}}\alpha^4
  +\cdots
\end{equation}



The renormalization equation for $\alpha'$ is
\begin{equation} \label{aprga}
u\frac{\mathrm{d}\alpha^{\prime}}{\mathrm{d}u}
=-\sum_{i=0}^{\infty} \beta_i^{\prime}{\alpha^{\prime}}^{i+2}.
\end{equation}
The primed beta function $\beta_i^{\prime}$ can be obtained
as such. Operating with $u\frac{\mathrm{d}}{\mathrm{d}u}$
on both sides of Eq.\ (\ref{apdef}), applying Eqs.\ (\ref{aprga}) and
(\ref{RGa}), and then comparing coefficients will give
$
\sum_{k=0}^i
\left[
\beta_k^{\prime}\bigsqcup_0^{k+2}a_{i-k}
-(k+1)a_k \beta_{i-k}
\right]=0,
$\
 namely, $\beta_i^{\prime}$ are given by the recursive relation
\begin{equation}
\beta_i^{\prime} 
=
  \sum_{k=0}^i
   (k+1)a_k \beta_{i-k}
   -\sum_{k=0}^{i-1}
   \beta_k^{\prime}\bigsqcup_0^{k+1}a_{i-k}.
\end{equation}
On application of Eqs.\ (\ref{a012exp}) and (\ref{a3exp}),
one immediately has the following explicit expressions
for the new beta function:
\begin{eqnarray}
\beta_0^{\prime}
&=& \beta_0=11/2-N_{\mathrm{f}}/3, \\
\beta_1^{\prime}
&=& \beta_1=51/4-(19/12)N_{\mathrm{f}}, \\
\beta_2^{\prime}
&=&
 \beta_2+a_2\beta_0
 -a_1(\beta_1+a_1\beta_0)
\nonumber\\
&=&
   \frac{2857}{64}
  -\frac{4549}{576}N_{\mathrm{f}}
  +\frac{79}{576}N_{\mathrm{f}}^2, \\
\beta_3^{\prime}
&=&
    \beta_3 +2a_3\beta_0
    -2a_1\beta_2+a_1^2\beta_1+4a_1^3\beta_0-6a_1a_2\beta_0
     \nonumber\\
&=&
     \beta_3^{(0)}+\beta_3^{(1)} N_{\mathrm{f}}
    +\beta_3^{(2)} N_{\mathrm{f}}^2
    +\frac{10085}{186624}N_{\mathrm{f}}^3, \\
 &\cdots&   \nonumber
\end{eqnarray}
with
$\beta_3^{(0)}=149753/768+(891/32)\zeta_3\approx 228.4606573$,
$\beta_3^{(1)}=2118091/82944+(2603291/55296)\zeta_3\approx 82.12826576$,
$\beta_3^{(2)}=-4027/1296-(194353/82944)\zeta_3\approx -5.923892810$.

It should be mentioned that a different expression for $\beta_2'$
was previously given in Ref.\ \cite{Marciano1984PRD29}.
The difference is caused by the fact that a wrong value for $C_2$
was quoted there \cite{Bernreuther1982NPB197}.

As an application of the general relation between $f_{i,j}$ and
the beta function, one can develop another expansion which converges
much faster. For this one can observe, more carefully, the standard
expansion
\begin{equation} \label{alfexpn}
\alpha=
 \sum_{i=0}^{\infty}
 \frac{\beta_0}{\beta_1}
 {L^*}^{i+1}
 \sum_{j=0}^i
 f_{i,j} \ln^jL
\equiv \sum_{i=0}^{\infty} J_i.
\end{equation}
Representing the terms in this expansion with the corresponding
coefficients $f_{i,j}$, all the terms can be
arranged in a matrix as
\begin{equation}
[f_{i,j}]
=\left[
  \begin{array}{ccccccc}
   1       & 0       & 0       & 0       & 0       &    \\
   f_{1,0} & 1       & 0       & 0       & 0       &    \\
   f_{2,0} & f_{2,1} & 1       & 0       & 0       &  \vdots\\
   f_{3,0} & f_{3,1} & f_{3,2} & 1       & 0       &    \\
   f_{4,0} & f_{4,1} & f_{4,2} & f_{4,3} & 1       &    \\
   \phantom{xxxx} &\phantom{xxxx} & \cdots
   & \phantom{xxxx} &\phantom{xxxx} & \phantom{xxx}
  \end{array}
 \right].
\end{equation}
The standard expansion corresponds to summing the terms row by row.
When one takes the expansion to a finite order, i.e., replacing
the $\infty$\ in Eq.\ (\ref{alfexpn}) with a positive integer $N$,
as has been done in the usual way,
then all the terms like $f_{j,j}\ (N<j<\infty)$ on the diagonal
and $f_{j+1,j}$ on the next to diagonal are missed, although these terms
are all known and have nothing to do with beta functions.
Generally, the terms $f_{j+k,j}$ for $0<j<\infty$ on the $k$th
next to diagonal involves only $\beta_{0\le l\le k-2}$.
But all the terms $f_{j+k,j}$ with $j>N$ are lost, though
no such terms are zero even when one sets all $\beta_{i>2}$ to
zero.

To include the contribution from the terms just mentioned,
we can consider to sum over diagonals,
which can be achieved by taking $i=j+k$ in Eq.\ (\ref{alfexpn}),
i.e.,
\begin{equation} \label{alfgen1}
\alpha=
\frac{\beta_0}{\beta_1}
\sum_{k=0}^{\infty}
\sum_{j=0}^{\infty}
f_{j+k,j} \left(L^*\ln L\right)^j
{L^*}^{k+1}
\equiv \sum_{k=0}^{\infty} I_k,
\end{equation}
where the expressions for $f_{j+k,j}$ can be obtained from
Eq.\ (\ref{figexpgen}):
\begin{eqnarray}
f_{j,j}
&=&
 1,  \ \
f_{j+1,j}
=
 \sum_{l=1}^j\left(\frac{j+1}{l}-1\right)+(j+1)C, \\
f_{j+2,j}
&=&\frac{\bar{\beta}_2}{2}(j+1)(j+2)
 +\left(C+\frac{1}{2}\right)
  \sum_{s=0}^{j-1} \frac{(s+1)(s+2)}{j-s}
 \nonumber\\
&&
 +\frac{1}{2}\sum_{s=0}^{j-2}\sum_{r=1}^{j-s-1}
    \frac{(s+1)(s+2)}{r(j-s-r)},
\ \cdots
\end{eqnarray}
From these expressions, we can give compact form to $I_k$:
\begin{eqnarray}
I_0
&=&
 \frac{L}{\beta_0}
 \sum_{j=0}^{\infty} (L^*\ln L)^j
=\frac{L/\beta_0}{1-L^*\ln L}
\equiv \beta_0 X,
\label{I0exp}\\
I_1
&=&
 \beta_0\beta_1
 X^2
 \left[
  C-\ln x
 \right], \\
I_2
&=&
 \beta_0\beta_1^2
 X^3
 \left[
  f_{2,0}-f_{2,1}\ln x +\ln^2x
 \right],
\label{I2exp}
\end{eqnarray}
where
$
x\equiv
1+(\beta_1/\beta_0^2)
\ln\ln(u/\Lambda)/\ln(u/\Lambda), \
\mbox{and}\
X\equiv
L^*/(\beta_1x)
=1/[\beta_0^2\ln(u/\Lambda)+\beta_1\ln\ln(u/\Lambda)].
$

\begin{figure}[htb]
\centering
\mbox{\includegraphics[width=8.0cm]{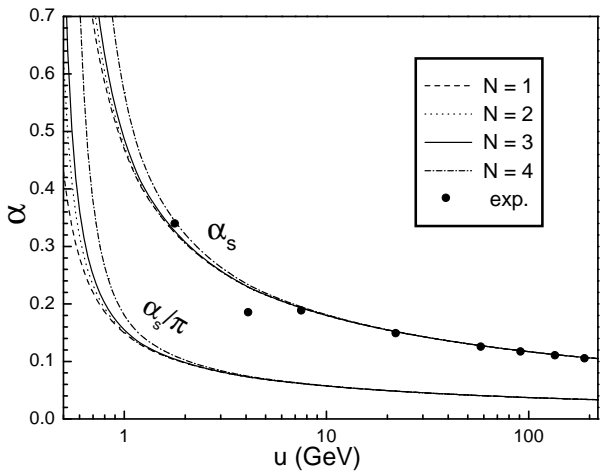}}
\caption{
 The QCD coupling as functions of the t'Hooft
 unit of mass. The dot, dash, solid, and dot-dash lines are,
 respectively, for the order $N$ from 1 to 4 with the scale $\Lambda$\
 indicated in Tab.\ 1. The full dots are the experimental data
 of, in increasing order of $\Lambda$,  $\tau$\ width,
 $\Upsilon$\ decay, deep inelastic scattering,
 e$^+$e$^-$ event shapes at 22 GeV from the JADE data,
 shapes at TRISTAN at 58 GeV, Z width, and e$^+$e$^-$
 event shapes at 135 and 189 GeV.
         }
 \label{aL}
\end{figure}

\begin{figure}[htb]
\centering
\mbox{\includegraphics[width=8.0cm]{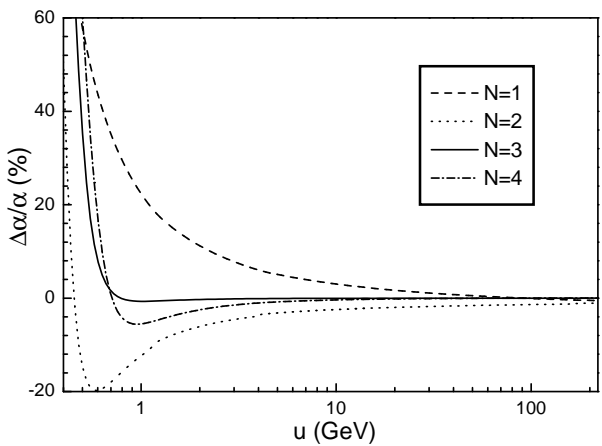}}
\caption{
 The relative difference between the new expansion in Eq.\ (\ref{alfgen2})
 and the conventional expansion in Eq.\ (\ref{alfexpn}).
         }
 \label{aL2}
\end{figure}

In Eq.\ (\ref{alfgen1}), there are an infinite number of logarithmic
terms like $\ln^j x$ which are included in $I_k$, even when one
takes the expansion to a finite order, say $\alpha=\sum_{k=0}^NI_k$.
The first several $I_k$s have been worked out to a closed form
in Eqs.\ (\ref{I0exp})--(\ref{I2exp}). The procedure is, to some extent,
similar to that in Ref.\ \cite{Bass2002PRD68} for the matching function.
Here we can, in fact, give the closed form for $I_k$ to arbitrary order:
\begin{equation} \label{alfgen2}
\alpha=
\sum_{k=0}^{\infty}
 \beta_0 \beta_1^k X^{k+1}
\sum_{l=0}^k
 (-1)^lf_{k,l}\ln^l x.
\end{equation}

In Eqs.\ (\ref{alfgen2}) and (\ref{alfexpn}), there are two arbitrary
constants: $\Lambda$ and $C$. Because the renormalization group
equation (\ref{aprga}) or (\ref{RGa}) is of the first order, only one
of them is independent. So we can arbitrarily take one of them,
while the other is determined by giving an initial condition.
It nearly becomes standard, nowadays, to take $C=0$ \cite{Bardeen1978PRD18},
which makes expressions somewhat simpler, and
$\alpha(m_{\mathrm{Z}})=0.1187/\pi$, where $M_{\mathrm{Z}}=91.1876$ GeV
is the mass of Z bosons. Setting $C=0$ requires distinct $\Lambda$ for
different effective flavor regimes, and we use $\Lambda_6$,  $\Lambda_5$,
$\Lambda_4$, and $\Lambda_3$ for $u>m_t$, $m_b<u<m_t$,
$m_c<u<m_b$, and $m_s<u<m_c$, respectively, where
the relevant quark masses are taken, in the present calculations, to be
$m_t=175$ GeV, $m_b=4.2$ GeV, $m_c=1.2$ GeV, and $m_s=100$ MeV.

In Fig.\ \ref{aL}, the coupling is shown as a function of the
renormalization point, calculated from Eq.\ (\ref{alfgen2})
with the infinity replaced by $N$, and the order $N$ from 1 to 4.
The same calculation has also been performed from the conventional
expansion in Eq.\ (\ref{alfexpn}). The relative difference between the
results from Eq.\ (\ref{alfgen2}) and Eq.\ (\ref{alfexpn}) is shown in
Fig.\ \ref{aL2}. It can be seen that, with decreasing $u$, the difference
becomes more and more significant.

\begin{table}[htb]
\centering
\caption{
 QCD renormalization group scale parameter $\Lambda$\ for the
 order from 1 to 4. For each $\Lambda_i$, the left column is
 for the new expansion in Eq.\ (\ref{alfgen2}) and the right column
 is for the conventional expansion in Eq.\ (\ref{alfexpn}).
        }
\begin{tabular}{|c|c|c|c|c|c|c|c|c|}\hline\hline
$\Lambda$ (MeV) & \multicolumn{2}{c|}{$\Lambda_6$}
  & \multicolumn{2}{c|}{$\Lambda_5$} & \multicolumn{2}{c|}{$\Lambda_4$}
  &\multicolumn{2}{c|}{$\Lambda_3$} \\ \hline
$N=1$& 88 & 45 & 208 & 91  & 286 & 124 & 325 & 147 \\ \hline
$N=2$& 91 & 95 & 217 & 235 & 303 & 335 & 350 & 377 \\ \hline
$N=3$& 90 & 90 & 214 & 212 & 298 & 294 & 340 & 334 \\ \hline
$N=4$& 88 & 86 & 209 & 207 & 295 & 291 & 344 & 342 \\ \hline\hline
\end{tabular}
\label{tabLam}
\end{table}

To compare the convergence speed of Eqs.\ (\ref{alfgen2}) and (\ref{alfexpn}),
all the $\Lambda_i$ ($i=3$--$6$) are listed in Tab.\ \ref{tabLam}.
There are two columns corresponding to each $\Lambda_i$,
the left column is for Eq.\ (\ref{alfgen2}) while the right column
is for Eq.\ (\ref{alfexpn}).
It is obvious that
the new expansion (\ref{alfgen2}) converges much faster than
the original expansion (\ref{alfexpn}). Even at the leading-order ($N=1$),
the corresponding $\Lambda_i$ for Eq.\ (\ref{alfgen2}) has nearly
approached to its value at order 4. So in practical applications,
it should be very accurate to calculate the coupling simply by
\begin{equation}
\alpha=
\frac{\beta_0}
     {\beta_0^2\ln(u/\Lambda)
      +\beta_1\ln\ln(u/\Lambda)
     }.
\end{equation}

In summary, the general relation between the standard expansion
coefficients and the beta function is carefully derived.
A matching-invariant coupling is given with the corresponding
beta function to four-loop level. A new expansion for
the coupling is then deduced, which is in principle more
accurate than the conventional expansion due to the inclusion of
an infinite number of logarithmic terms in a closed form.

\appendix

\section{square cup operator}

The square cup operator, $\bigsqcup_m^k$, is defined so that
it meets
\begin{equation}
\left(
 \sum_{i=m}^{\infty} a_i x^i
\right)^k
=\sum_{i=0}^{\infty}
 \left(
 \bigsqcup_m^k a_i
 \right)
 x^i.
\end{equation}
Obviously one has
$
\bigsqcup_m^0 a_i =\delta_{i,0}, \
\bigsqcup_m^1 a_{i<m}=0, \
\bigsqcup_m^1 a_{i\ge m} =a_i.
$\
 And for $k\ge 2$,
it is also not difficult to give a general explicit expression
\begin{equation}
\bigsqcup_m^k a_i
= \left(
   \prod_{s=1}^{k-1}
   \sum_{p_s=m}^{i-(k-s)m-\varsigma_s^p}
  \right)
   a_{i-\varsigma_k^p} \prod_{r=1}^{k-1} a_{p_r}
\end{equation}
where
%
\begin{equation}
\varsigma_s^p
 \equiv
\left\{
\begin{array}{ll}
\sum_{t=1}^{s-1}p_t & \mathrm{if}\ s>1 \\
         0          & \mathrm{otherwise}
\end{array}
\right..
\end{equation}
The meaning of $\varsigma_k^p$ is similar to this.
Here are several special simple cases:
\begin{equation}
\bigsqcup_m^k a_0=\delta_{k,0}, \ \
\bigsqcup_m^k a_{i\le km}=0, \ \
\bigsqcup_m^k a_{km}=a_m^k, \ \
\end{equation}

The two-dimensional extension of the square cup operator,
$\bigsqcup_{m,n}^k$, is defined by
\begin{equation}
\left(
 \sum_{i=m}^{\infty}\sum_{j=n}^i
 f_{i,j} x^i y^j
\right)^k
=\sum_{i=0}^{\infty}\sum_{j=0}^i
 \left(
 \bigsqcup_{m,n}^k f_{i,j}
 \right)
 x^i y^j.
\end{equation}
Similarly, one has
$
\bigsqcup_{m,n}^0 f_{i,j}=\delta_{i,0}\delta_{j,0}, \
\bigsqcup_{m,n}^1 f_{i<m,j<n}=0, \
\bigsqcup_{m,n}^1 f_{i\ge m,j\ge n}=f_{i,j}.
$\
And for $k\ge 2$, we have
\begin{equation}
\bigsqcup_{m,n}^k f_{i,j}
=\left(
 \prod_{s=1}^{k-1} \sum_{p_s=m}^{i-(k-s)m-\varsigma^p_s}
               \sum_{q_s=\sigma}^{p_s^*}
 \right)
 f_{i-\varsigma^p_k,j-\varsigma^q_k}
 \prod_{r=1}^{k-1} f_{p_r,q_r}
\end{equation}
where
\vspace{-0.5cm}
\begin{eqnarray}
p_s^*
&\equiv& \mathrm{min} \left[p_s,j-(k-s)n-\varsigma_s^q\right], \\
\sigma
 &\equiv& \mathrm{max}\left(n,\sum_{t=1}^{s}p_t -\varsigma_s^q-i+j\right).
\end{eqnarray}
Here are special examples:
\begin{equation}
\bigsqcup_{m,n}^k f_{0,0}=\delta_{k,0}, \
\bigsqcup_{m,n}^k f_{i<km,j<kn}=0, \
\bigsqcup_{m,n}^k f_{km,kn}=f_{m,n}^k.
\end{equation}

\section*{Acknowledgements}

The author acknowledges support from
DOE (DF-FC02-94ER40818),
NSFC (10375074, 90203004, and 19905011),
FONDECYT (3010059 and 1010976),
CAS (E-26), and SEM (B-122).
He also acknowledges hospitality at
the MIT center for theoretical physics,
the INFN laboratori nazionali del sud,
the PUC faculdad de physica,
and the IN2P3 institut des sciences nucl\'{e}aires.

\end{document}